\documentstyle[12pt]{article}
\newcommand{\gsim}{\mathrel{\hbox{\rlap{\lower.55ex \hbox {$\sim$}}
                   \kern-.3em \raise.4ex \hbox{$>$}}}}
\newcommand{\lsim}{\mathrel{\hbox{\rlap{\lower.55ex \hbox {$\sim$}}
                   \kern-.3em \raise.4ex \hbox{$<$}}}}
\newcommand {\vectwo}[2] {\left(\begin{array}{c}#1\\#2\end{array}\right)}

\newcommand {\beq}    {\begin{equation}}
\newcommand {\eeq}    {\end{equation}}

\newcommand {\thetaw} {\theta_{w}}

\newcommand {\twoi}   {2\, i}

\newcommand {\ncls}{non-contractible loops}
\newcommand {\ncs} {non-contractible sphere}

\newcommand {\Ncss}{Non-contractible spheres}
\newcommand {\ncss}{non-contractible spheres}

\newcommand {\sm}  {standard model}

\newcommand {\sph}   {spha\-le\-ron}
\newcommand {\sphs}  {sphalerons}

\newcommand {\cs} {configuration space}
\newcommand {\ew} {electroweak}

\newcommand {\Sstar} {S$^{\star }$}

\newcommand {\ewsmodel} {electroweak standard model}

\newcommand {\art}      {article }

\begin{document}
\begin{titlepage}
\hspace*{\fill}
NIKHEF-H/94-02
\newline
\hspace*{\fill}
February 2, 1994
\begin{center}
	\vspace{2\baselineskip}
	{\Large \bf A new perspective on \ew ~strings}\\
	\vspace{1\baselineskip}
\renewcommand{\thefootnote}{\fnsymbol{footnote}}
	{\large
	 F. R. Klinkhamer \\
	 CHEAF/NIKHEF--H \\ Postbus 41882 \\
	 NL--1009 DB Amsterdam \\ The Netherlands \\
	\vspace{1\baselineskip}
         and\\
	\vspace{1\baselineskip}
         P. Olesen\\
	 The Niels Bohr Institute\\
         University of Copenhagen\\ Blegdamsvej 17\\
         DK--2100 Copenhagen\\Denmark\\
	}
\renewcommand{\thefootnote}{\arabic{footnote}}
\setcounter{footnote}{0}
	\vspace{3\baselineskip}
	{\bf Abstract} \\
\end{center}
{\small \par
The vortex solution ($Z$-string) of the \ew ~interactions can be interpreted
as the 2-dimensional \sph ~at the top of a \ncs.
The same holds for another type of solution, the $W$-string.
}
\end{titlepage}

\section{Introduction}

It has been known for a long time that the vortex solution \cite{NO73}
of the Abelian Higgs model may be embedded \cite{N77} in the \ew ~\sm,
where it excites the $Z^{0}$ and Higgs fields.
Recently, the stability properties of this \ew ~vortex solution ($Z$-string)
have been investigated numerically \cite{JPV93}.
The $Z$-string solution turns out to be metastable
for values of the weak mixing angle $\thetaw$ very close to $\pi/2$
and small enough Higgs mass.
For the physical mixing angle $\thetaw \sim \pi/6\,$, however,
the solution is unstable, most likely for all values of the Higgs mass.
The existence and potential stabilty of the $Z$-string for
$\thetaw \rightarrow \pi/2$ are well understood \cite{H92,V92}.
In this \art we start from $\thetaw =0$ and
``explain'' the existence and {\em instability\/} of the solution,
by showing analytically that, for $\thetaw \leq \pi/4$ at least,
the $Z$-string is the 2-dimensional \sph
{}~at the top of a \ncs ~in \cs.
Recall that, generally speaking, a sphaleron
denotes a static, but unstable, finite energy solution of the classical
field equations, where the theory may be defined in any number of spatial
dimensions.

The outline of this \art is as follows.
In Sect. 2 we establish our notation.
In Sect. 3 we construct a \ncs ~of 2-dimensional configurations
of the fields and in Sect. 4 we show that, for small enough values of
$\thetaw$,
its maximum corresponds to the $Z$-string solution.
We also construct in Sects. 3 and 4
another, simpler  \ncs, which is related to a different
solution, the $W$-string \cite{BVB94},  with larger energy per unit length
than the $Z$-string.
The main thrust of this paper is analytical, but, for completeness, we
present in Sect. 4 some numerical results for both types of solutions
and their corresponding \ncss.
In Sect. 5, finally, we summarize our main observations
and present a simple picture of \cs.

\section{Notation}

We consider static classical fields in the bosonic sector of the \ewsmodel,
namely the $SU(2)$ gauge field
$W_{m} \equiv W_{m}^{a}\: \tau^{a} \,$, the $U(1)$ hypercharge
gauge field $B_{m}$ and the complex Higgs doublet field $\Phi$.
The energy functional of these fields is given by
\beq
 E = \int_{{\rm R}^3} d^{3}x
 \left[ \frac{1}{4 \,g^{2}}\; \left( W_{m n}^{a} \right)^{2} +
        \frac{1}{4 \,g^{\prime \,2}}\; \left( B_{m n}     \right)^{2} +
        |D_{m} \Phi|^{2} +
         \lambda \left( |\Phi|^{2} - v^{2}/2 \right)^{2} \: \right] ,
\label{eq:E}
\eeq
with the following definitions for the field strengths and
covariant derivatives
\begin{eqnarray*}
W_{mn} &\equiv& W_{mn}^{a}\: \tau^{a}  \equiv
  \partial _{m} W_{n} - \partial _{n} W_{m} + [W_{m},W_{n}]\\
B_{m n} &\equiv&  \partial _{m} B_{n} - \partial _{n} B_{m}\\
D_{m}\Phi&\equiv&\left( \partial_{m} + W_{m}^{a}\; \frac{\sigma^{a}}{\twoi }
                         + B_{m}\; \frac{1}{\twoi } \right) \Phi \: ,
\label{eq:definitions}
\end{eqnarray*}
where the indices run over the values $1,\, 2,\,3$, and
$\tau^{a} \equiv \sigma^{a}/(\twoi)$  are the $SU(2)$ generators,
with $\sigma^{a}$ the standard Pauli matrices.
The semiclassical masses of the $W^{\pm}$ and $Z^{0}$ vector bosons are
$M_{W}=\frac{1}{2}\, g\, v$
and $M_{Z} = M_{W} / \cos \thetaw$, with the weak mixing angle
$\thetaw$ defined as $\tan \thetaw \equiv g' / g$.
The mass of the physical Higgs scalar is
$M_{H}=\sqrt{8\,\lambda/g^{2}}\, M_{W}$.

In this \art we are interested in axisymmetric
field configurations which are constant along the symmetry axis.
We use  the standard cylindrical coordinates $\rho$, $\phi$ and $z$,
defined in terms of the cartesian coordinates by
$(x_{1},x_{2},x_{3}) \equiv$ ($\rho \cos \phi\,$, $ \rho \sin \phi\,$, $z$).
Making all distances dimensionless ($\tilde{\rho} \equiv
\rho M_{Z}\/$ etc. , and dropping the tildes afterwards) we then write for the
total energy of these fields
\beq
E \equiv
    \frac{2 \pi v \, \cos\thetaw}{g} \;
    \int_{-\infty}^{\infty}dz \: \epsilon \; =
    \frac{2 \pi v \, \cos\thetaw}{g} \;
    \int_{-\infty}^{\infty}dz \; \int_{0}^{\infty} d\rho\,\rho
    \;\; e(\rho) \; ,
\label{eq:e}
\eeq
with $e$ the dimensionless energy density and $\epsilon$
the dimensionless energy per unit length (string tension).  The total
energy $E$ is infinite for the configurations we consider and, instead,
we focus on the 2-dimensional energy $\epsilon$, which should be finite.
Specifically, we investigate the topology of the space of finite
$\epsilon$ configurations and in the next section we will take two
slices through this infinitely dimensional \cs.
\footnote{
We consider in this \art the minimal \sm, but we expect our topological
considerations to have more general validity.
          }

\section{\Ncss}

In this section we present two \ncss ~of 2-dimensional axi\-sym\-me\-tric
field configurations,
i. e. configurations that depend on $\rho$ and that have
vanishing $z$-components of the gauge fields.
They differ in that the first \ncs ~(NCS) has $SU(2)$ gauge fields only,
whereas the second NCS excites both the $SU(2)$ and $U(1)$ gauge fields.
We will call the first \ncs ~
the $W$-NCS and the second the $Z$-NCS, because the first has at the top
$W^{\pm}$ gauge fields whereas the second has a $Z^{0}$ gauge field instead.

\subsection{$W$-NCS}

The NCS is parametrized by  the square $\mu,\nu \in [-\pi,+\pi]$, with the
boundary $ |\mu|=\pi$ or $|\nu|= \pi$ corresponding to the classical vacuum.
Writing $[\mu\nu] \equiv \max( |\mu|,|\nu|)$ and using differential forms,
the field configurations of the NCS are
\newcommand {\ftil}    {\tilde{f}}
\newcommand {\htil}    {\tilde{h}}
\newcommand {\ktil}    {\tilde{k}}
\newcommand {\Util}    {\tilde{U}}
\newcommand {\Omegatil}{\tilde{\Omega}}
\begin{eqnarray}
\pi/2 \leq [\mu\nu] \leq \pi & :
                         &W   =  0  \nonumber\\
                         &&B  =  0  \nonumber\\
                         &&\Phi=  \left( 1 - (1-\htil) \sin [\mu\nu]\,\right)
                        \: \frac{v}{\sqrt{2}} \: \vectwo{0}{1}  \nonumber\\
                         &&  \nonumber\\
0 \leq[\mu\nu]<\pi/2 & :
  &W   = -\ftil\:\Omegatil\: {\rm d}\Util\: \Util^{-1}\:\Omegatil^{-1}
           \nonumber\\
  &&B  = 0 \nonumber\\
  &&\Phi=\htil\:\frac{v}{\sqrt{2}}\:\Omegatil\:\Util \vectwo{0}{1},
\label{eq:WNCS}
\end{eqnarray}
with the $SU(2)$ matrices
\beq
\Util(\mu,\nu,\phi) =
\left( \begin{array}{l}
       \:\sin \mu                    \\
       \:\cos \mu \:\sin \nu           \\
       \:\cos \mu \:\cos \nu \:\sin \phi \\
       \:\cos \mu \:\cos \nu \:\cos \phi
       \end{array}
\right)
\cdot
\left( \begin{array}{c}
       -i \sigma_{2} \\
       -i \sigma_{3} \\
        1            \\
       -i \sigma_{1}
       \end{array}
\right)
\label{eq:Util}
\eeq
and
\beq
\Omegatil = \Util(\mu,\nu,0)^{-1} =
\sin \mu \;             i \sigma_{2} +
\cos \mu \: \sin \nu \; i \sigma_{3} +
\cos \mu \: \cos \nu \; i \sigma_{1} \; .
\label{eq:Omegatil}
\eeq
The matrix $\Util$ gives  a mapping of the 3-sphere into
$SU(2)$ with unit winding number
and $\Omegatil$ implements a global gauge transformation of the fields.
The axial functions $\ftil(\rho)$ and $\htil(\rho)$ in
(\ref{eq:WNCS}) have the following boundary conditions
\begin{eqnarray}
\lim_{ \rho \rightarrow \infty} \ftil,\htil &=& 1  \nonumber\\
\ftil(0)=\htil(0)                           &=& 0  \; ,
\label{eq:Wbcs}
\end{eqnarray}
which guarantee, respectively, vanishing energy density at infinity
and regularity of the fields at the core.
Note that the 2-dimensional fields of the NCS (\ref{eq:WNCS}) are in the radial
gauge ($W_{\rho} = B_{\rho} =0$) and that, in general,
the remaining global gauge freedom
is fixed by the value of the Higgs field $\Phi$ at the point $\phi =0$, $\rho
\rightarrow \infty$.

The dimensionless energy density (\ref{eq:e})
becomes for $0 \leq[\mu\nu] < \pi/2$
\begin{eqnarray}
e &=& \cos^{2}\mu \: \cos^{2}\nu \:
      \left[\:
      \cos^{-2}\thetaw \: \rho^{-2}\:(\partial_{\rho} \,\ftil)^{2} +
      \rho^{-2}\:\htil^{2}\: \left( 1-\ftil \right)^{2} \:
      \right]  + \nonumber\\
  & & (\partial_{\rho} \,\htil)^{2} +
      \frac{1}{4} \left( \frac{M_{ H}}{M_{ Z}} \right)^{2}
      \left( \htil^{2} -1 \right)^{2}
\label{eq:Weballoon}
\end{eqnarray}
and for $\pi/2 \leq [\mu\nu] \leq \pi$
\beq
e =
    (\partial_{\rho} \,\ktil)^{2} +
    \frac{1}{4} \left( \frac{M_{ H}}{M_{ Z}} \right)^{2}
    \left( \ktil^{2} -1 \right)^{2} \; ,
\label{eq:Wecord}
\eeq
with $\ktil \equiv  1 - (1-\htil) \sin [\mu\nu] $.
Manifestly, the maximum of the 2-dimensional energy $\epsilon(\mu,\nu)$ occurs
at $\mu = \nu =0$.
\footnote {
           In order to have a {\em manifest\/} maximum the
ansatz (\ref{eq:WNCS}) is split into a part that ``rotates'' the gauge fields
(taking the Higgs along for the ride) and a part that ties the Higgs field to
its vacuum value, rather than performing the two operations together.
           }
The maximum energy $\epsilon(0,0)$ can be minimized
by solving the variational equations from (\ref{eq:e}, \ref{eq:Weballoon}), and
the resulting functions are denoted by
$\overline{\ftil}$ and $\overline{\htil}$.
This procedure is called a minimax procedure.

\subsection{$Z$-NCS}

The construction of the second NCS starts from a ``rotated'' version of the
$SU(2)$ matrix (\ref{eq:Util}), which, at $\mu = \nu =0$ in particular,
allows for the  excitation of a $U(1)$ gauge field.
Using the same notation as before ($\tau^{a} \equiv \sigma^{a}/(\twoi)$),
the field configurations of this NCS are
\begin{eqnarray}
\pi/2 \leq [\mu\nu] \leq \pi & :
                         &W   =  0  \nonumber\\
                         &&B  =  0  \nonumber\\
                         &&\Phi=  \left( 1 - (1-h) \sin [\mu\nu]\,\right)
                        \: \frac{v}{\sqrt{2}} \: \vectwo{0}{1}  \nonumber\\
                         &&  \nonumber\\
0 \leq[\mu\nu] <   \pi/2 & :
                         &W   = -f\:G^{a} \, \tau^{a}
                                  \nonumber\\
                         &&B  =  f\: \sin^{2}\thetaw \: F^{3}  \nonumber\\
                         &&\Phi=h\:\frac{v}{\sqrt{2}}\:\Omega\:U \vectwo{0}{1},
\label{eq:ZNCS}
\end{eqnarray}
with the following Lie algebra valued 1-forms
\begin{eqnarray}
F^{a} \, \tau^{a} & \equiv & U^{-1} \, {\rm d}U \nonumber \\
G^{a} \, \tau^{a} & \equiv & \Omega \: U \: \left[
F^{1} \, \tau^{1} +  F^{2} \,\tau^{2} + \cos^{2}\thetaw \: F^{3} \,\tau^{3}
                             \right] \: U^{-1} \, \Omega^{-1}
\label{eq:FGdefinitions}
\end{eqnarray}
and $SU(2)$ matrices
\begin{eqnarray}
U(\mu,\nu,\phi) &=&
\left( \begin{array}{l}
       \:\sin \mu                    \\
       \:\cos \mu \:\sin \nu           \\
       \:\cos \mu \:\cos \nu \:\sin \phi \\
       \:\cos \mu \:\cos \nu \:\cos \phi
       \end{array}
\right)
\cdot
\left( \begin{array}{c}
       -i \sigma_{1} \\
       -i \sigma_{2} \\
       -i \sigma_{3} \\
        1
       \end{array}
\right)\\
\label{eq:U}
&& \nonumber \\
\Omega &=& U(\mu,\nu,0)^{-1} \; .
\label{eq:Omega}
\end{eqnarray}
The axial functions $f(\rho)$ and $h(\rho)$ in (\ref{eq:ZNCS})
have the usual boundary
conditions
\begin{eqnarray}
\lim_{ \rho \rightarrow \infty} f,h &=& 1  \nonumber\\
f(0)=h(0)                           &=& 0  \; ,
\label{eq:Zbcs}
\end{eqnarray}
in order to have smooth fields with finite 2-dimensional energy $\epsilon$.

The dimensionless energy density (\ref{eq:e})
becomes for $0 \leq[\mu\nu] < \pi/2$
\begin{eqnarray}
e &=& \cos^{2}\mu \: \cos^{2}\nu \:
      \left(\: 1 -  \cos^{2}\mu \: \cos^{2}\nu  \: \sin^{2}\thetaw\:\right)\:
      \cos^{-2}\thetaw\:\left[ \:\rho^{-2}\:(\partial_{\rho}\,f)^{2}\:\right]
      + \nonumber\\
  & & \cos^{2}\mu \: \cos^{2}\nu \;
      \left[ \: \rho^{-2}\:h^{2}\: \left( 1-f \right)^{2} \: \right] +
      (\partial_{\rho} \,h)^{2} +
      \frac{1}{4} \left( \frac{M_{ H}}{M_{ Z}} \right)^{2}
      \left( h^{2} -1 \right)^{2}
\label{eq:Zeballoon}
\end{eqnarray}
and for $\pi/2 \leq [\mu\nu] \leq \pi$
\beq
e =
    (\partial_{\rho} \,k)^{2} +
    \frac{1}{4} \left( \frac{M_{ H}}{M_{ Z}} \right)^{2}
    \left( k^{2} -1 \right)^{2} \; ,
\label{eq:Zecord}
\eeq
with $k \equiv  1 - (1-h) \sin [\mu\nu] $.
For small enough values of $\thetaw$,
the maximum of the 2-dimensional energy $\epsilon(\mu,\nu)$ occurs
at $\mu = \nu =0$.  Again, the minimax procedure is to solve the
variational equations from (\ref{eq:e}, \ref{eq:Zeballoon}) at
$\mu = \nu =0$,
and the resulting functions are denoted by $\overline{f}$ and $\overline{h}$.
In the next section we will investigate the optimal maximum
configuration of this NCS further.

\section{Sphalerons}

By now it is well known that the topology of the configuration
space of the \ew ~\sm ~gives rise to non-trivial
classical solutions. Specifically, the existence of \ncls ~and \ncss ~of
3-dimensional configurations of the bosonic fields leads to, respectively, the
\sph ~S \cite{KM84} and the \sph ~\Sstar ~\cite{K93}.
In the same way we may expect the \ncss ~of 2-dimensional configurations,
as constructed in the previous section, to be related to non-trivial
string-like classical solutions. In this Section we will show that
these solutions correspond to the known $W$- and $Z$-string solutions.
Furthermore, we will demonstrate that
these solutions are at the top of their respective \ncss, so that they
are really \sphs ~(for the $Z$-string this holds provided $\thetaw$ is
small enough).
\newcommand {\fW}{f_{W}}
\newcommand {\hW}{h_{W}}
\newcommand {\fZ}{f_{Z}}
\newcommand {\hZ}{h_{Z}}

\subsection{$W$-string}

The maximum configuration of  the $W$-NCS  (\ref{eq:WNCS})
reproduces, up to a global gauge transformation,
to the so-called $W$-string solution \cite{BVB94}, which excites the
$W^{\pm} \equiv (W^{1} \mp i W^{2})/ \sqrt{2}$ gauge fields.
More precisely, the minimax procedure over the $W$-NCS gives the
$W$-string solution with axial functions
$\fW =  \overline{\ftil}$ and $\hW = \overline{\htil}$.
{}From the energy density (\ref{eq:Weballoon}) we then find
the energy  profile over the $W$-NCS for $0 \leq[\mu\nu] < \pi/2$
\begin{eqnarray}
\epsilon(\mu,\nu) &=& \epsilon_{W} \: \left[ \:
  \cos^{2}\mu \: \cos^{2}\nu \:\left(\tilde{a}+\tilde{b} \right) +
  \tilde{c} + \tilde{d}\:   \right] \;,
\label{eq:epsilonWNCS}
\end{eqnarray}
in terms of the following integrals
\begin{eqnarray}
\tilde{a} + \tilde{b} &\equiv&
            \epsilon_{W}^{-1} \:\int_{0}^{\infty} d\rho\:\rho
            \left[\: \cos^{-2}\thetaw\:\rho^{-2}\:(\partial_{\rho}\,\fW)^{2}+
            \rho^{-2}\:\hW^{2}\: \left( 1-\fW \right)^{2}\:\right] \nonumber\\
\tilde{c} + \tilde{d} &\equiv&
            \epsilon_{W}^{-1} \;\int_{0}^{\infty} d\rho\, \rho \:
            \left[\: (\partial_{\rho} \, \hW)^{2} +
            1 / 4 \, \left( M_{ H}/ M_{ Z} \right)^{2}
            \left( \hW^{2} -1 \right)^{2} \:\right]  \; ,
\label{eq:actilde}
\end{eqnarray}
which add up to 1.
The energy $\epsilon(\mu,\nu)$ for $\pi/2 \leq [\mu\nu] \leq \pi$
decreases monotonically with $[\mu\nu]$, as follows by inspection
of (\ref{eq:Wecord}).
In table 1 we give some numerical results for the energy $\epsilon_{W}$
and the integrals (\ref{eq:actilde}).
The relation $\epsilon_{W}(\sin^{2}\thetaw = 3/4\,;  M_{H}/M_{Z}) =
              \epsilon_{W}(\sin^{2}\thetaw = 0\,; 2\,M_{H}/M_{Z})$
displayed in this table follows directly from the scaling properties of
$\epsilon_{W}$ as given by (\ref{eq:actilde}).
The dimensionless energy $\epsilon_{W}$ is, in fact, a function of
$\cos^{-1}\thetaw \, M_{H}/M_{Z} = M_{H}/M_{W}$ only.
In short, we have  $\epsilon_{W} = \epsilon_{W}(M_{H}/M_{W})$.

{}From the energy profile (\ref{eq:epsilonWNCS}) we conclude immediately that
the $W$-string
is the 2-dimensional \sph ~of a \ncs, with at least two negative modes.
This holds for arbitrary values of $\thetaw$ and $M_{H}/M_{Z}$.
We turn now to the other type of string solution, where the situation is
more complicated.

\subsection{$Z$-string}

First, recall that the fields of the \ew ~vortex solution ($Z$-string)
\cite{NO73,N77} can be written in the following form
\begin{eqnarray}
W    &=& - \cos^{2}\thetaw \: \fZ \: {\rm d}V\: V^{-1} \nonumber\\
B    &=& - \tan^{2}\thetaw \: W^{3}  \nonumber\\
\Phi &=& \hZ \:\frac{v}{\sqrt{2}}\:V \vectwo{0}{1},
\label{eq:NO}
\end{eqnarray}
where the $SU(2)$ matrix $V$ is defined in terms of the matrix
$U(\mu,\nu,\phi)$ of eq. (\ref{eq:U}) by
\beq
V \equiv U(0,0,\phi)
\eeq
and the functions $\fZ$ and $\hZ$ are the solutions of the
field equations with boundary conditions (\ref{eq:Zbcs}).
The corresponding energy density
\beq
e_{Z} =
      \rho^{-2}\:(\partial_{\rho} \, \fZ)^{2} +
      \rho^{-2}\:\hZ^{2}\: \left( 1-\fZ \right)^{2}+
      (\partial_{\rho} \, \hZ)^{2} +
      \frac{1}{4} \left( \frac{M_{ H}}{M_{ Z}} \right)^{2}
      \left( \hZ^{2} -1 \right)^{2}
\label{eq:eNO}
\eeq
is independent of the value of the weak mixing angle $\thetaw$.
This energy density gives the dimensionless $Z$-string energy
$\epsilon_{Z} = \epsilon_{Z}(M_{H}/M_{Z})$, which is related to the
$W$-string energy in the following simple way
\beq
\epsilon_{Z}(M_{H}/M_{Z}) = \epsilon_{W}(M_{H}/M_{W}) \; .
\label{eq:epsZepsW}
\eeq
The relation (\ref{eq:epsZepsW}) is not entirely unexpected,
considering the fields involved in both types of string solutions.
Their stability properties, however, may be different, depending on
the value of the weak mixing angle $\thetaw$.

For $\thetaw = 0$ we observe from (\ref{eq:Zeballoon}) that
the 2-dimensional energy  $\epsilon(\mu,\nu)$ of the $Z$-NCS has
a manifest maximum at $\mu=\nu=0$ and the minimax procedure over the NCS
(\ref{eq:ZNCS}) leads precisely to the $Z$-string (\ref{eq:NO}),
with $\fZ = \overline{f}$ and $\hZ = \overline{h}$.
The $Z$-string solution, for $\thetaw =0$,
has thus at least two negative modes.
By continuity, we expect the same conclusion to hold for small
enough values of $\thetaw$. This can be investigated with the
general ansatz (\ref{eq:ZNCS}) for the $Z$-NCS.

{}From the energy density (\ref{eq:Zeballoon}) we find,
using the axial functions of the $Z$-string $f = \fZ $ and $h = \hZ$,
the energy  profile over the $Z$-NCS for $0 \leq[\mu\nu] < \pi/2$
\begin{eqnarray}
\epsilon(\mu,\nu) &=& \epsilon_{Z} \: \left[ \:
      \cos^{2}\mu \: \cos^{2}\nu \:
      \left(\: 1 -  \cos^{2}\mu \: \cos^{2}\nu  \: \sin^{2}\thetaw\:\right)\:
      \cos^{-2}\thetaw\:a \right. + \nonumber\\
  & & \left. \cos^{2}\mu \: \cos^{2}\nu \; \: b + c + d \: \right] \;,
\label{eq:epsilonZNCS}
\end{eqnarray}
in terms of the following integrals
\begin{eqnarray}
a  &\equiv& \epsilon_{Z}^{-1} \:\int_{0}^{\infty} d\rho\:
            \rho^{-1}\:\left(\partial_{\rho} \, \fZ\right)^{2}  \nonumber \\
b  &\equiv& \epsilon_{Z}^{-1} \;\int_{0}^{\infty} d\rho\:
            \rho^{-1}\:\hZ^{2}\: \left( 1-\fZ \right)^{2} \nonumber \\
c  &\equiv& \epsilon_{Z}^{-1} \;\int_{0}^{\infty} d\rho\, \rho \:
            \left(\partial_{\rho} \, \hZ\right)^{2} \nonumber \\
d  &\equiv& \epsilon_{Z}^{-1} \;\int_{0}^{\infty} d\rho\, \rho \:
            1 / 4 \, \left( M_{ H}/ M_{ Z} \right)^{2}
            \left( \hZ^{2} -1 \right)^{2}  \; ,
\label{eq:abcd}
\end{eqnarray}
which add up to 1.
The energy $\epsilon(\mu,\nu)$ for $\pi/2 \leq [\mu\nu] \leq \pi$
decreases monotonically with $[\mu\nu]$, as follows by inspection
of (\ref{eq:Zecord}).
Expanding (\ref{eq:epsilonZNCS}) around the $Z$-string at
$\mu=\nu=0$ gives
\beq
\epsilon = \epsilon_{Z} \: \left[ \:
           1 + (\mu^{2} + \nu^{2})
           \left( \tan^{2}\thetaw - 1 - b / a \right) a +
           {\rm O}(\mu^{4},\, \nu^{4},\, \mu^{2}\nu^{2})  \: \right] \;.
\label{eq:NOinstability}
\eeq
Since $a$ and $b$ are both positive definite we have in
(\ref{eq:NOinstability}) manifest
instability as long as $ \tan^{2}\thetaw \leq 1$.
For these parameter values we can also
verify from (\ref{eq:epsilonZNCS}) that the local maximum at $\mu=\nu=0$
is in fact a global maximum of the NCS.
The $Z$-string, for $\thetaw \leq \pi/4$ and arbitrary Higgs mass,
has thus at least two negative modes.
This applies, in particular, to the physical mixing angle
$\thetaw \sim \pi/6$.
In short, the $Z$-string solution  of the \ew ~interactions
is the 2-dimensional \sph ~of a \ncs.
\footnote{
          Note that this result is different from the heuristic
          connection  between the $Z$-string and the 3-dimensional
          \sph ~S  as discussed in ref. \cite{BVB94}.
         }

Numerical results allow
us to extend the range of instability beyond $\thetaw = \pi/4$
($\sin^{2}\thetaw = 1/2$).
In table 2  we present some numerical data for the integrals
(\ref{eq:abcd}), together with the resulting maximum value
$\thetaw^{\rm NCS}$ for instability, as follows from (\ref{eq:NOinstability}).
For $\thetaw < \thetaw^{\rm NCS}$ the $Z$-string  solution has two
negative modes, made explicit by the fields (\ref{eq:ZNCS}) of the $Z$-NCS.
This range of instability is consistent with the results
of ref. \cite{JPV93}. Finally, we remark that the results of
table 2 may be used to map out the complete energy profile
(\ref{eq:epsilonZNCS}) of the \ncs ~through the $Z$-string,
as a function of both $\thetaw$ and $M_{H}/M_{Z}$.
For the case of perturbative stability \cite{JPV93} of the $Z$-string,
we find an energy barrier for decay towards the vacuum,
which is of the order of the energy $\epsilon_{W}$ of the $W$-string.

\section{Summary}

We have constructed in this \art two \ncss, which are related to certain
string-like classical solutions, the so-called $W$- and $Z$-strings.
It is perhaps surprising that there are {\em two}
\ncss, since the mapping ${\rm S}^{3} \rightarrow SU(2)$ with unit winding
number is essentially unique.
If there were only $SU(2)$ gauge fields the \ncs ~would indeed be unique
(up to global gauge transformations), but this is not the case
when the Higgs and $U(1)$ gauge fields are included.
For the second \ncs, in particular, the interplay of these fields
is quite subtle in keeping the  2-dimensional energy $\epsilon$ down,
for small enough values of the weak mixing angle $\thetaw$.

We present in fig. 1 a schematic picture of \cs.
As far as the $W$-string solution is concerned, the situation is quite
simple. The $W$-string is, for all values of $\thetaw$, the \sph ~of a
\ncs ~(\ref{eq:WNCS}). The solution has thus two (or more) negative modes.
As far as the $Z$-string  solution is concerned, the situation is more
complicated.  Starting from $\thetaw =0$, also the $Z$-string is the \sph ~of
a \ncs ~(\ref{eq:ZNCS}). For increasing values of $\thetaw$
the ``balloon'' of the non-contractible sphe\-re tilts out of the $SU(2)$ plane
and the $Z$-string moves away in configuration space,
in order to keep its energy $\epsilon$ constant, and ultimately,
for $\thetaw = \pi/2$, its $SU(2)$ gauge field content vanishes altogether.
As $\thetaw$ approaches $\pi/2$ the negative modes of the $Z$-string
turn over and the solution becomes metastable.
However, for the physical mixing angle $\thetaw \sim \pi/6$
the $Z$-string solution is still
a genuine sphaleron, with two (or more) negative modes.

\vspace{1\baselineskip}
One of us (F. R. K.) thanks T. Vachaspati for interesting him in
e\-lec\-tro\-weak strings, the Aspen Center for Physics for hospitality
and the Leids Kerkhoven--Bosscha Fonds for a travel grant.
We both thank E. Faverey for drawing the figure.
\vspace{1\baselineskip}

\newcommand {\eZ}    {\epsilon_{Z}}
\newcommand {\stwNCS}{\sin^{2}\theta_{w}^{\rm NCS}}
\newcommand {\stw}   {\sin^{2}\theta_{w}}
\begin{table}[p]
\begin{center}
\begin{tabular}{lccccccc}
\hline
\hline
           & \multicolumn{7}{c}{$M_{H}/M_{Z}$}    \\
           &$1/8$  &$1/4$  &$1/2$  &$1$    &$2$    &$4$    &$8$   \\
\hline
$\stw=0$   &$0.47$ &$0.59$ &$0.76$ &$1.00$ &$1.34$ &$1.79$ &$2.35$ \\
           &$0.34$ &$0.39$ &$0.44$ &$0.50$ &$0.56$ &$0.63$ &$0.69$ \\
           &$0.66$ &$0.61$ &$0.56$ &$0.50$ &$0.44$ &$0.37$ &$0.31$ \\
           &       &       &       &       &       &       &       \\
$\stw=1/4$ &$0.49$ &$0.62$ &$0.80$ &$1.06$ &$1.43$ &$1.90$ &$2.47$ \\
           &$0.35$ &$0.40$ &$0.46$ &$0.52$ &$0.58$ &$0.65$ &$0.71$ \\
           &$0.65$ &$0.60$ &$0.54$ &$0.48$ &$0.42$ &$0.35$ &$0.29$ \\
           &       &       &       &       &       &       &       \\
$\stw=1/2$ &$0.52$ &$0.67$ &$0.87$ &$1.16$ &$1.55$ &$2.06$ &$2.65$ \\
           &$0.36$ &$0.42$ &$0.47$ &$0.54$ &$0.60$ &$0.66$ &$0.72$ \\
           &$0.64$ &$0.58$ &$0.53$ &$0.46$ &$0.40$ &$0.34$ &$0.28$ \\
           &       &       &       &       &       &       &       \\
$\stw=3/4$ &$0.59$ &$0.76$ &$1.00$ &$1.34$ &$1.79$ &$2.35$ &$2.97$ \\
           &$0.39$ &$0.44$ &$0.50$ &$0.57$ &$0.63$ &$0.70$ &$0.75$ \\
           &$0.61$ &$0.56$ &$0.50$ &$0.43$ &$0.37$ &$0.30$ &$0.25$ \\
\hline
\hline
\end{tabular}
\end{center}
\caption[]{ \protect
Numerical results for the $W$-string solution :
upper, middle and lower entries stand for, respectively, the dimensionless
energy $\epsilon_{W}$, the integrals $\tilde{a}+\tilde{b}$ and
the integrals $\tilde{c}+\tilde{d}$.
}
\end{table}

\begin{table}[p]
\begin{center}
\begin{tabular}{lccccccc}
\hline
\hline
           & \multicolumn{7}{c}{$M_{H}/M_{Z}$}    \\
           &$1/8$  &$1/4$  &$1/2$  &$1$    &$2$    &$4$    &$8$   \\
\hline
$\eZ$      &$0.47$ &$0.59$ &$0.76$ &$1.00$ &$1.34$ &$1.79$ &$2.35$ \\
$a$        &$0.16$ &$0.18$ &$0.20$ &$0.21$ &$0.21$ &$0.20$ &$0.18$ \\
$b$        &$0.19$ &$0.22$ &$0.25$ &$0.29$ &$0.35$ &$0.43$ &$0.52$ \\
$c$        &$0.50$ &$0.43$ &$0.36$ &$0.29$ &$0.22$ &$0.16$ &$0.12$ \\
$d$        &$0.15$ &$0.17$ &$0.19$ &$0.21$ &$0.22$ &$0.21$ &$0.18$ \\
$\stwNCS$  &$0.69$ &$0.69$ &$0.69$ &$0.70$ &$0.73$ &$0.76$ &$0.79$ \\
\hline
\hline
\end{tabular}
\end{center}
\caption[]{ \protect
Numerical results for the $Z$-string solution :
the dimensionless energy $\eZ$, the integrals $a$--$d$ and the
critical mixing angle $\theta_{w}^{\rm NCS}$.
}
\end{table}

\newpage
\section*{Figure caption }
{\bf Fig. 1 } Sketch of \cs, with vertically the
energy $\epsilon$ of the 2-dimensional configurations considered.
The directions in \cs ~of
the $SU(2)$ and $U(1)$ gauge fields are indicated symbolically.
The mass ratio $M_{H}/M_{Z}$ is considered fixed.
For mixing angle $\thetaw =0$
the $SU(2)$ \ncs ~(\ref{eq:ZNCS}) is shown with one dimension suppressed :
the ``balloon'', with parameter values $0 \leq[\mu\nu] < \pi/2$,
is attached by a ``cord'', with $\pi/2 \leq[\mu\nu]\leq \pi$, to the
vacuum V.  The maximum of this \ncs ~coincides with
the $Z$-string solution. Further away in \cs ~there
is another solution ($W$-string), which is degenerate in energy.
For $\thetaw$ increasing towards its physical value
($ \sim \pi/6$)  the \ncs ~tilts in the $U(1)$ direction and
the $Z$-string moves away,
keeping its energy $\epsilon$ constant, whereas the maximum of the
pure $SU(2)$ \ncs ~(\ref{eq:WNCS}) gets a larger energy and corresponds to
another solution ($W$-string).
The picture is schematic in that these two \ncss ~are not really
``parallel'', not even approximately.

\end{document}